\documentclass[twocolumn,9pt]{article} 

\usepackage[square,numbers,sort&compress,comma]{natbib}

\usepackage[dvipsnames]{xcolor}

\usepackage{amsmath}
\usepackage{amssymb}
\usepackage{caption}
\usepackage{graphicx}
\usepackage{latexsym}
\usepackage{times}
\usepackage[pagewise]{lineno}
\usepackage{hyperref}
\usepackage{chemmacros}
\usepackage{chemfig}
\usepackage{chemformula}
\usepackage{siunitx}
\usepackage{float}
\usepackage{enumitem}

\topmargin - 12pt 
\newcommand{\mrm}[1]{\mathrm{#1}}

\newcommand{\upd}{\mathrm{d}}                           
\newcommand{\der}[2]{\frac{\upd #1}{\upd #2}}           

\chemsetup{
  formula = mhchem ,
  modules  = {reactions}
}

\renewenvironment{abstract}%
              {
               \small
               {\bfseries \abstractname}
               \par
               \vspace{10pt}
              }

\renewcommand\abstractname{Abstract}

\newcommand{\nomenclature}
              [1]
              {
               \bgroup
               \flushleft
               \small\bf
               #1
               \par
               \egroup
              }

\renewcommand{\section}
              [1]
              {
               \bgroup
               \flushleft
               \small\bf
               \refstepcounter{section}
               \arabic{section}. #1
               \par
               \egroup
              }

\renewcommand{\subsection}
              [1]
              {
               \bgroup
               \flushleft
               \small\em
               \refstepcounter{subsection}
               \arabic{section}.
               \arabic{subsection}. #1
               \par
               \egroup
              }

\renewcommand{\subsubsection}
              [1]
              {
               \bgroup
               \flushleft
               \small\em
               \refstepcounter{subsubsection}
               \arabic{section}.
               \arabic{subsection}.
               \arabic{subsubsection}. #1
               \par
               \egroup
              }

  \newcommand{\acknowledgement}
              [1]
              {
               \bgroup
               \flushleft
               \small\bf
               #1
               \par
               \egroup
              }

  \newcommand{\sectionbib}
              [1]
              {
               \bgroup
               \flushleft
               \small\bf
               #1
               \par
               \egroup
              }

\begin{document}



\small
\baselineskip 10pt

\setcounter{page}{1}
\title{\LARGE \bf LES of iron-powder combustion in a jet-in-hot-coflow burner --
                    Insights on flame structure and ignition characteristics}

\author{{\large Shyam Hemamalini$^{a,b}$, XiaoCheng Mi$^{a,b,*}$}\\[10pt]
        {\footnotesize \em $^a$Department of Mechanical Engineering, Eindhoven University of Technology, Eindhoven, Netherlands}\\[-5pt]
        {\footnotesize \em $^b$Eindhoven Institute for Renewable Energy Systems, Eindhoven, Netherlands}\\[-5pt]}

\date{}

\twocolumn[\begin{@twocolumnfalse}
\maketitle
\rule{\textwidth}{0.5pt}
\vspace{-5pt}

\begin{abstract}
Micron-sized iron powders are a rapidly advancing novel energy storage technology. In order to improve the design of real-world iron-powder combustors, adequate understanding of the ignition behavior in such settings is necessary. In this work, a jet-in-hot-coflow (JHC) burner designed by Hameete~\textit{et al}. \cite{hameetethesis, hameete2024aerosol} to test ignition of iron particles in lab-scale turbulent flames is modeled numerically using LES and Lagrangian point-particles. The JHC burner is simulated in two different modes--an open flame and an enclosed flame--similar to the experimental reference. Two iron-oxidation-rate models--the first-order model \cite{Hazenberg} and oxide-layer model \cite{Mi2022}--are used to examine the effect on capturing the ignition behavior. For the radiative heat transfer between the two phases, a simplified Stefan-Boltzmann approximation model and the P1 model are considered, similar to Ramaekers \emph{et al.} \cite{ramaekers2025}. Analysis of flame structure indicates ignition in the circumference of the jet, aided by the break-up of the coflow. At higher $T_\mrm{coflow}$, ignition onset and oxidation completion is earlier prior to jet break-up. Minimum coflow temperature for particle ignition with the oxide layer model is $T_\mrm{coflow} = \SI{1125}{\kelvin}$ with complete oxidation at $T_\mrm{coflow} = \SI{1250}{\kelvin}$, and for the first-order model at $\SI{800}{\kelvin}$ and $\SI{900}{\kelvin}$, respectively. Both of these results do not match the experimental results of Hameete \cite{hameetethesis}. Oxidation degree is predictably higher for enclosed flames. For the chosen particle distribution, the P1 model exhibits higher radiative heat loss and results in a slightly lower oxidation degree. Analysis on particle ensembles with partial oxidation shows that the overall oxidation degree at a sufficient height above the nozzle reflects particle ignition probability. Further analysis in regards to particle size shows ignition failure is more prevalent in larger particles.
\end{abstract}

\vspace{10pt}

{\bf Novelty and significance statement}

This study provides one of the first numerical insights into turbulent iron powder flames. Ignition (or lack of) is a critical problem facing industrial iron powder combustors. This work compares commonly used numerical models that describe the oxidation of iron particles on predicting the ignition behavior in such flames. The numerical framework and the results from this study are vital in continuing the understanding on iron cloud ignition, and to design efficient combustors of the future.

\vspace{5pt}
\parbox{1.0\textwidth}{\footnotesize {\em Keywords:} Metal Fuels; Turbulent Iron Combustion; Large Eddy Simulations; Jet-in-hot-coflow}
\rule{\textwidth}{0.5pt}
*Corresponding author. Email: x.c.mi@tue.nl
\vspace{5pt}
\end{@twocolumnfalse}] 

\section{Introduction\label{sec:introduction}} \addvspace{10pt}

Micron-sized iron powders are a novel energy carrier that has seen rapid scientific and industrial advancement over the past two decades. Pilot reactors of up to 1 MW of operational power have been demonstrated \cite{niekvanrooijthesis, van20260}. However, in such large-scale industrial reactors, a prominent problem observed is incomplete oxidation and lack of ignition. To improve the design of such combustors, an in-depth understanding of the ignition process is necessary. While many studies \cite{chang2026experimental,cen2025detailed, Ning2024,Abdallah_2024, CEN2026} have focused on the ignition of single, isolated particles, the complexities of a realistic large-scale reactor remain underexplored. The flow in a typical large-scale combustor is turbulent, and iron powder flames, owing to the heterogeneous and non-volatile nature of the iron particles, are essentially particle-laden turbulent flows. There have been several studies on the modeling of turbulent iron combustion \cite{Hemamalini2024,Luu2024,Thaeter2024,thater2026interaction,steffens2025exploring,HEMAMALINI2026}, although the majority focus on canonical flow scenarios such as mixing layers and homogeneous isotropic turbulence. Steffens \emph{et al.} \cite{steffens2025exploring} were the first to perform a combined experimental and numerical study of a large-scale methane-assisted iron combustor. The study showed complex flow phenomena such as particle clustering, and good agreement on the flow features was reported; however, the phenomenon of particle ignition was not the focus.

Experimentally, Hameete \emph{et al.} \cite{hameete2024aerosol} developed a lab-scale turbulent burner that employs a jet-in-hot-coflow with micron-sized iron particles injected into the turbulent coflow. A crucial result from Hameete \cite{hameetethesis} shows that the oxidation degree at various coflow temperatures correlates with the trend of \textit{probability} of particle ignition determined via single-particle experiments with the same particles by Muhammad \emph{et al.} \cite{Abdallah_2024}. However, lower ignition temperatures than those from the single-particle results were observed, and it was hypothesized that this is the effect of the collective heating of the particles. This finding provides statistical insight into cloud ignition in a turbulent setting. However, a deeper in situ analysis of the burner is unavailable at the time of writing. 

In the work of Hameete \textit{et al}. \cite{hameete2024aerosol,hameetethesis}, a turbulent flame was stabilized in two modes--an open flame and an enclosed flame--and only a posteriori statistics of particle oxidation were reported due to the lack of available optical diagnostic techniques to measure such flames. Very recently, Hebel \emph{et al.} \cite{hebel2026ignition} conducted a series of experiments wherein a particle jet was ejected into a hot coflow using a prior experimental setup from the same group \cite{eitel2015novel} on turbulent non-premixed hydrocarbon flames. In this work, however, the flame is optically accessible, and the authors used Mie scattering and luminosity imaging to measure velocity and temperature profiles at critical locations.

From a numerical perspective, there are two commonly used frameworks to model iron particle oxidation: the first-order model formulated by Hazenberg \emph{et al.} \cite{Hazenberg} and the oxide-layer model formulated by Mi \emph{et al.} \cite{Mi2022}. These models differ significantly in their description of solid-state kinetic reaction rates. The ignition temperatures for isolated particles predicted by the oxide-layer model better agree with experimental measurements~\cite{chang2026experimental,cen2025detailed, Ning2024,Abdallah_2024}. Recently, Mi \cite{mi2025theoretical} has shown that the oxide-layer model can effectively capture the hindering effect of the growth of the oxide-layer on the ignition of the iron particles. 

This work explores computationally the same jet-in-hot-coflow burner developed by Hameete \emph{et al.}\cite{hameete2024aerosol}. The main objectives of this work are:
\begin{enumerate}[wide, labelindent=0pt,itemsep=0pt]
    \item to compare the two iron oxidation models--the first-order model and the oxide-layer model--in capturing the ignition behavior.
    \item to evaluate the oxidation degree of the iron powder flame under two operational modes -- an open flame and a closed flame.
    \item to model radiative heat transfer through a simplified Stefan-Boltzmann approximation model and the P1 radiation model, and to compare the corresponding ignition characteristics.
\end{enumerate}

Additionally, since \textit{in situ} data from such particle-jet-in-hot-coflow is non-existent at the time of writing, the flame structure from the simulations is also analyzed and interpreted. 

\section{Methodology\label{sec:methodology}} \addvspace{10pt}

\subsection{Geometry and experimental data\label{subsec:geometry}} 
\begin{figure}[h]
    \centering
    \includegraphics[width=0.65\linewidth]{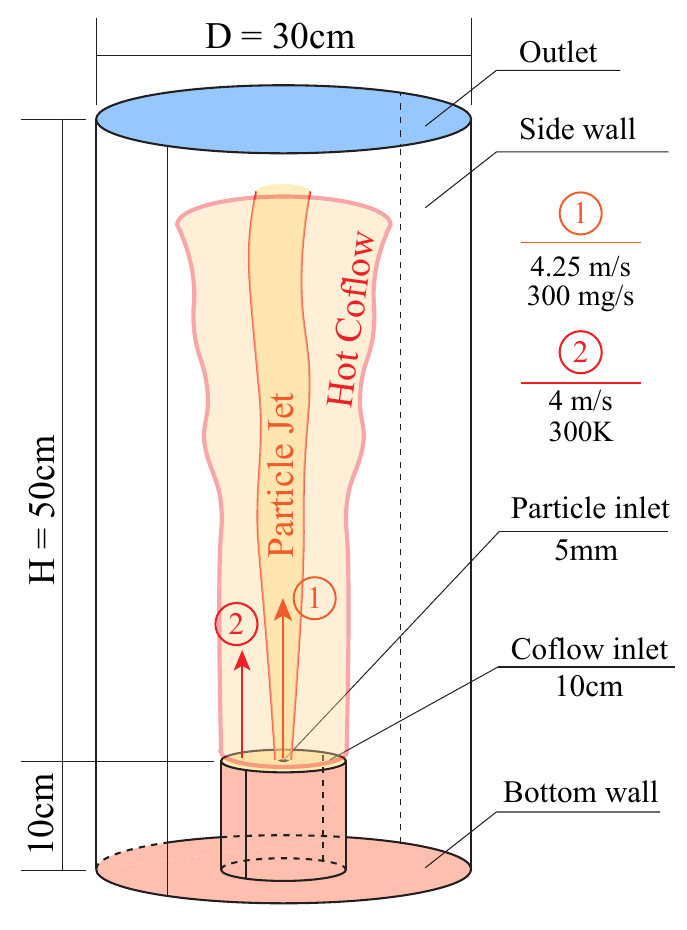}
    \caption{Jet-in-hot-coflow geometry and flow parameters considered in this work}
    \label{fig:domain}
\end{figure}

\addvspace{10pt}
The geometry is modeled with reference to the jet-in-hot-coflow experimental setup of Hameete \emph{et al.}\cite{hameete2024aerosol}. In this work, both modes of operation experimentally conducted by Hameete \emph{et al.}--an open flame and an enclosed flame with a cylindrical enclosure made of white brick--are considered. Figure \ref{fig:domain} shows the geometrical configuration, where a cylindrical domain of diameter $\SI{30}{\centi\meter}$ and height $\SI{60}{\centi\meter}$ is modeled. The hot coflow is modeled as a circular surface of diameter $\SI{10}{\centi\meter}$, elevated from the bottom surface of the domain by a height of $\SI{10}{\centi\meter}$. The particle injection inlet is also modeled as a circular surface in the inlet with a diameter of $\SI{5}{\milli\meter}$.

The coflow velocity is $U_\mrm{coflow} = \SI{4}{\meter/\second}$ in the $+z$-direction, and the particle jet velocity is $U_\mrm{p} = \SI{4.25}{\meter/\second}$. The bulk Reynolds number of the coflow is computed to be $\mrm{Re} \approx 3000$, indicating a transitional regime of turbulence. The Richardson number of the jet is computed assuming ambient air at $T_\infty = \SI{300}{\kelvin}$ and a maximum coflow temperature of $T_\mrm{coflow} = \SI{1300}{\kelvin}$ to be $\mrm{Ri} \approx 0.204$. The particles are injected from the injection tube with a mass flow rate of $\dot{m}_\mrm{p} = \SI{300}{\milli\gram/\second}$. The reported Sauter diameter of the particle distribution is $D_\mrm{3,2} = \SI{46.93}{\micro\meter}$. The flow in the injection tube is laminar with an estimated Reynolds number of $\mrm{Re} \approx 1150$. 

\subsection{Governing Equations \label{subsec:equations}} \addvspace{10pt}

To model the particle-jet-in-hot-coflow, a two-way coupled Eulerian-Lagrangian framework is used to conduct Large Eddy Simulations (LES) with the open-source solver \textit{OpenFOAM}. For this purpose, a custom solver incorporating the oxidation-rate model of $\mrm{Fe}$ particles has been implemented by Ramaekers \emph{et al.} \cite{ramaekers2025}, to which we refer the reader for an extensive description. In the following text, the salient features of the framework are presented.

\subsubsection{Carrier-phase model\label{subsec:gasphase}} \addvspace{10pt}

The gas-phase is modeled using a three-dimensional Eulerian grid. The compressible form of the Navier-Stokes equations pertaining to the conservation of mass, momentum, energy, and species mass fraction is solved. Since the simulations are of type LES, the Favre-filtered form of the governing equations appear as:
\begin{align}
    &\frac{\partial \bar{\rho}_\mrm{g}}{\partial t} + \nabla \cdot (\bar{\rho}_\mrm{g} \tilde{\mathbf{u}}_\mrm{g}) = \bar{S}_{\text{M}} \\
    &\frac{\partial \bar{\rho}_\mrm{g} \tilde{\mathbf{u}}_\mrm{g}}{\partial t} + \nabla \cdot (\bar{\rho}_\mrm{g} \tilde{\mathbf{u}}_\mrm{g} \tilde{\mathbf{u}}_\mrm{g}) = -\nabla \bar{p} + \bar{\rho}_\mrm{g} \mathbf{g}  \\[-0.5em]
    &\phantom{\frac{\partial \bar{\rho}_\mrm{g} \tilde{\mathbf{u}}_\mrm{g}}{\partial t} + \nabla \cdot (\bar{\rho}_\mrm{g} \tilde{\mathbf{u}}_\mrm{g} \tilde{\mathbf{u}}_\mrm{g})} + \nabla \cdot (\bar{\boldsymbol{\tau}} - \boldsymbol{\tau}^{\text{sgs}}) + \bar{\mathbf{S}}_{\text{F}} \notag\\
    &\frac{\partial \bar{\rho}_\mrm{g} \tilde{h}_{\text{s}}}{\partial t} + \nabla \cdot (\bar{\rho}_\mrm{g} \tilde{h}_{\text{s}} \tilde{\mathbf{u}}_\mrm{g}) = \frac{D\bar{p}}{Dt}  \\[-0.5em]
    &\phantom{\frac{\partial \bar{\rho}_\mrm{g} \tilde{h}_{\text{s}}}{\partial t} + \bar{\rho}_\mrm{g} \tilde{h}_{\text{s}}}+ \nabla \cdot \left( \frac{\lambda}{c_p} \nabla \tilde{h}_{\text{s}} - \mathbf{q}^{\text{sgs}} \right) + \bar{S}_{\text{H}} \notag\\
    &\frac{\partial \bar{\rho}_\mrm{g} \tilde{Y}_k}{\partial t} + \nabla \cdot (\bar{\rho}_\mrm{g} \tilde{Y}_k \tilde{\mathbf{u}}_\mrm{g}) = \nabla \cdot (\bar{\rho}_\mrm{g} D_k^{\text{eff}} \nabla \tilde{Y}_k - \mathbf{J}_k^{\text{sgs}}) \notag \\[-0.75em]
    &\phantom{\frac{\partial \bar{\rho}_\mrm{g} \tilde{Y}_k}{\partial t} + \nabla \cdot (\bar{\rho}_\mrm{g} \tilde{Y}_k \tilde{\mathbf{u}}_\mrm{g})}+ \bar{S}_{Y,k}
\end{align}\\[-6pt]
where $\bar{S}_{\text{M}}$, $\bar{S}_{\text{F}}$, $\bar{S}_{\text{H}}$, and $\bar{S}_{\text{Y}}$ are the corresponding Lagrangian source terms of mass, momentum, enthalpy, and species, respectively, $D_k^{\mathrm{eff}}$ the effective diffusivity, and the other terms with their usual notation. In this work, only $\mrm{O_2}$ and $\mrm{N_2}$ are considered as the gas-phase species. No gas-phase reactions are considered.

The subgrid scale stresses and fluxes $\boldsymbol{\tau}^{\text{sgs}}$, $\boldsymbol{q}^{\text{sgs}}$, and $\boldsymbol{J}^{\text{sgs}}$ are closed with the dynamic $k$-equation subgrid scale model \cite{kim1995new}, which solves for the subgrid scale kinetic energy $\tilde{k}_\text{sgs}$ using dynamically computed coefficients $C_k$ and $C_\epsilon$, and adapts well to transitional turbulence as well as wall-bounded flows.

\subsubsection{Particle model\label{subsec:particle}} \addvspace{10pt} 

The particles are modeled as Lagrangian spherical point-particles and are two-way coupled with the gas-phase Eulerian grid through mass $\bar{S}_{\text{M}}$, momentum $\bar{S}_{\text{F}}$, heat transfer $\bar{S}_{\text{H}}$, and oxygen consumption $\bar{S}_{\mathrm{Y_\mrm{O_2}}}$. The particle drag is modeled using the particle transport equation as:
\begin{align}
    \frac{\text{d} \mathbf{u}_{\text{p}}}{\text{d} t} = \frac{3}{4} \frac{\bar{\rho}_{\text{g}} C_{\text{D}}}{\rho_{\text{p}} d_{\text{p}}} \cdot |\tilde{\mathbf{u}}_{\text{g}} - \mathbf{u}_{\text{p}}| (\tilde{\mathbf{u}}_{\text{g}} - \mathbf{u}_{\text{p}})
\end{align}\\[-6pt]
where the drag-coefficient $C_\mrm{D}$ is given by the Schiller-Naumann drag correlation \cite{schiller1933drag} as:
\begin{equation}
    C_{\text{D}} = \frac{24}{\text{Re}_\mrm{p}} \left(1 + 0.15 \text{Re}_\mrm{p}^{0.687}\right)
\end{equation}\\[-6pt]
with $\mrm{Re}_\mrm{p}$ being the particle Reynolds number computed with the slip velocity between the particle and the local gas field. The Stokes number of the particles at a flow timescale in the SGS scale computed at the minimum cell size close to the injection nozzle evaluates to $\mrm{St}_\mrm{SGS,25\mu m} \approx 2.76$ and $\mrm{St}_\mrm{SGS,65\mu m} \approx 18.65$. The particles do not respond well to fluctuations smaller than the filter width, and hence, only the filtered gas velocity $\tilde{\mathbf{u}}_\mrm{g}$ is used in the calculation of particle drag.The particle model includes convective and evaporative heat transfer with the carrier phase, as in Ramaekers \emph{et al.} \cite{ramaekers2025} with the particle temperature solved explicitly from the enthalpy equation after the addition of convective, radiative, and evaporative terms as:
\begin{align}
    \der{H_\mrm{p}}{t} &= -\frac{h_\mrm{O_2}}{W_\mrm{O_2}} \dot{m}_\mrm{O_2} - \dot{q}_\mrm{conv} - \dot{q}_\mrm{rad} - \dot{q}_\mrm{evap} \\
    H_\mrm{p} &= \frac{m_\mrm{Fe}}{W_\mrm{Fe}} h_\mrm{Fe}(T_\mrm{p}) + \frac{m_\mrm{FeO}}{W_\mrm{FeO}} h_\mrm{FeO}(T_\mrm{p})
\end{align}
where $H_\mrm{p}$ is the particle enthalpy, $h_\mrm{O_2}$ is the specific enthalpy of $\mrm{O_2}$ at $T_\mrm{p}$, $\dot{q}_\mrm{conv}$ is the convective heat flux computed using the Ranz-Marshall correlation, 
and $\dot{q}_\mrm{rad}$ and $\dot{q}_\mrm{evap}$ are the radiative and evaporative fluxes, respectively.

The two widely-used particle reaction models are considered: the first-order model and the oxide-layer model. Only a single-stage reaction from Fe to FeO is considered. The implementation of the first-order model is based on Hazenberg \emph{et al.} \cite{Hazenberg}, where the $\mrm{O_2}$ mass consumption $\dot{m}_\mrm{O_2}$ by particles of size $d_\mrm{p}$ is approximated using the Damk\"ohler number $\mrm{Da}^*$ as:
\begin{equation}
    \dot{m}_\mrm{O_2} = \rho_\mrm{p}Y_\mrm{O_2}A_\mrm{d}k_\mrm{d}\mrm{Da^*} \label{eq:dam}
\end{equation}
where $\rho_\mrm{p}$ is the particle density, $A_\mrm{d} = A_\mrm{p}$ the particle surface area, and $\mrm{Da^*} = A_\mrm{r}k_\mrm{r}/(A_\mrm{r}k_\mrm{r} + A_\mrm{d}k_\mrm{d})$ the normalized Damk\"ohler number. The diffusive reaction rate is given as $k_\mrm{d} = \mrm{Sh} D_\mrm{O_2}/ d_\mrm{p}$, where $\mrm{Sh}$ and $D_\mrm{O_2}$ represent the Sherwood number and the binary diffusion coefficient of oxygen in the film layer, respectively. The surface reaction rate is given by an Arrhenius-type relation as $k_\mrm{r} = k_\infty \exp(-T_\mrm{a}/T_\mrm{p})$. In the current work, $k_\infty = \SI{7.5e6}{\meter/\second}$ and $T_\mrm{a}=\SI{14.4e3}{\kelvin}$ are used, which are the same coefficients as those used by several prior studies. In the current implementation, the first-order model exhibits a single-particle isobaric ignition temperature of $T_\mrm{ign}\approx\SI{775}{\kelvin}$ for the particle size ranges considered in this study, where $T_\mrm{ign}$ is the minimum gas temperature required for thermal runaway.

The implementation of the oxide-layer model is based on the framework by Mi \emph{et al.} \cite{Mi2022}, where $\dot{m}_\mrm{O_2}$ is modeled as a switch-type rate based on the interplay between the solid-state $\mrm{Fe}^+$ diffusion through the oxide layer $\dot{m}_\mrm{O_2,R}$ and the diffusion of $\mrm{O}_2$ from the bulk gas to the particle surface through the film layer $\dot{m}_\mrm{O_2,D}$ as:
\begin{equation}
    \begin{split}
        \dot{m}_\mrm{O_2} &= \dot{m}_\mrm{O_2,R}\quad\: \mrm{if}\ \dot{m}_\mrm{O_2,R}<\dot{m}_\mrm{O_2,D,max} \\
        &= \dot{m}_\mrm{O_2,D,max} \quad \mrm{otherwise} 
    \end{split}
\end{equation}
where $\dot{m}_\mrm{O_2,D,max} = A_\mrm{p}k_\mrm{d}\rho_\mrm{O_2,f}$ with $\rho_\mrm{O_2,f}$ the density of oxygen gas in the film layer, and 
\begin{equation}
    \dot{m}_\mrm{O_2,R} = s_\mrm{FeO} \cdot\frac{\rho_\mrm{FeO} A_\mrm{p} k_\mrm{0}}{X_\mrm{FeO}} \exp \left( \frac{-T_\mrm{a}}{T_\mrm{p}}\right)
\end{equation}
where $s_\mrm{FeO}$ is the stoichiometric ratio, $k_\mrm{0}$ the pre-exponential factor, and $X_\mrm{FeO}$ the oxide-layer thickness. In the current implementation, $k_0=\SI{2.67e-4}{\meter^2/\second}$ and $T_\mrm{a}=\SI{20319}{\kelvin}$ are used, and a single-particle isobaric ignition temperature of $T_\mrm{ign}\approx\SI{1070}{\kelvin}$ for the considered particle size ranges is observed. 

\subsubsection{Radiative heat transfer\label{subsec:radiation}} \addvspace{10pt}

In this work, two types of radiative heat transfer models are considered and compared in the simulations: the simplified particle-to-local-gas radiative exchange given by the Stefan-Boltzmann approximation and the P1 model using an Eulerian incident radiation field $G$. The P1 model, unlike more complex Discrete Ordinates Method (DOM) models, is computationally efficient and inexpensive, and more accurate on a global scale than the simplified approximation. 

The former model considers the radiative heat transfer from the particle $q_\mrm{rad,p}$ as follows:
\begin{equation}
    q_\mrm{rad,p,SB} = A_\mrm{p} \epsilon_\mrm{p} \sigma_\mrm{SB} \left(T_\mrm{p}^4 - \tilde{T}_\mrm{g}^4\right)
\end{equation}
where $\epsilon_\mrm{p}$ is the particle emissivity, assumed to be $\epsilon_\mrm{p} = 0.7$ in this work \cite{ramaekers2025},the Stefan-Boltzmann constant $\sigma_\mrm{SB} = 5.67\times10^{-8}\,\SI{}{\watt\meter^{-2}\kelvin^{-4}}$, $\tilde{T}_\mrm{g}$ the local gas temperature at the particle location. Alternatively, the P1 model for radiative heat transfer considers an incident radiation field $G$, solved with the carrier-phase governing equations, to which each particle adds or subtracts its contribution depending on the following equation:
\begin{equation}
    q_\mrm{rad,p,P1} = \epsilon_\mrm{p} A_\mrm{p} \left( \frac{G}{4} - \sigma_\mrm{SB} T_\mrm{p}^4 \right) \label{eq:p1}
\end{equation}
The well-known limitation of the P1 model is the over-estimation of radiative heat loss by the particles if the domain exhibits open boundaries and the media is optically thin \cite{modest2021radiative}.  

\subsection{Domain and boundary conditions} \addvspace{10pt}

\begin{figure}[h]
    \centering
    \includegraphics[width=\linewidth]{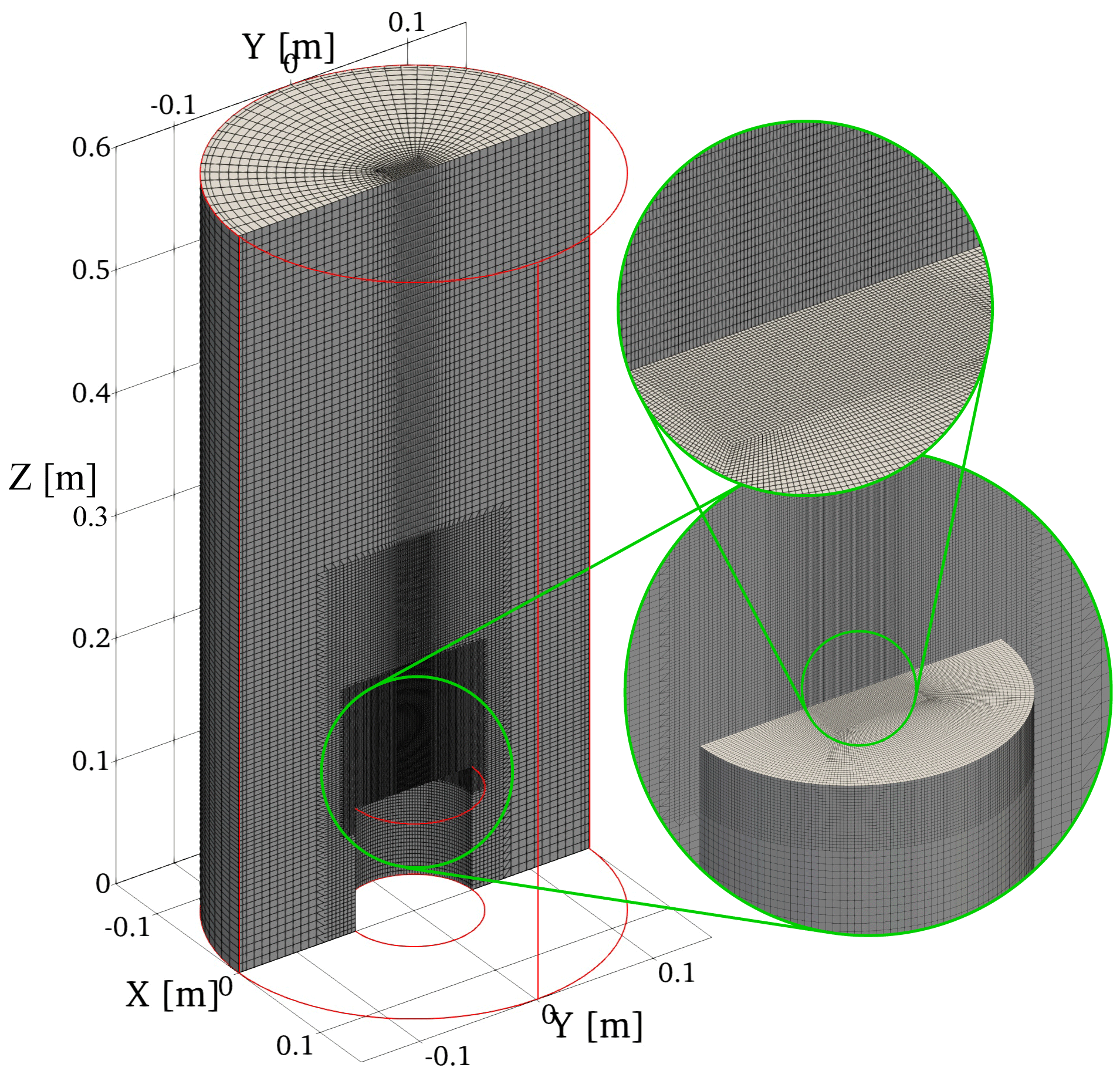}
    \caption{The hexahedral butterfly mesh used in the simulations of enclosed flame. Insets show a zoomed in view of the coflow inlet and particle injection nozzle.}
    \label{fig:mesh}
\end{figure}

The domain presented in Section \ref{subsec:geometry} is modeled as a cylindrical graded butterfly mesh consisting of 3,114,880 hexahedral cells for the simulations of the enclosed domain, as shown in Figure \ref{fig:mesh}, with a minimum cell width of $\Delta x = \SI{0.4375}{\milli\meter}$ at the particle injection nozzle and $\Delta x = \SI{0.875}{\milli\meter}$ at the edge of the coflow inlet. The minimum $y^+$ at the side and bottom walls is $\mrm{min}(y^+) \approx 0.15$, and the mean $y^+ \approx 1.74$ indicates adequate resolution for the chosen LES model. Evaluation of Pope's criterion is made to ensure sufficient turbulence resolution at critical regions (included in the supplementary material). Standard wall functions for $k$ and $\nu_\mrm{t}$ are implemented to capture near-wall flow effects. For each simulation, the flow is initialized and stabilized for 24 flow-through times of the coflow before particle injection. Gravity and buoyancy effects are not considered due to the low $\mrm{Ri}$ as calculated in Section \ref{subsec:geometry}.

The outlet has been modeled as a non-reflecting outflow boundary with an ambient pressure of $p_\infty = \SI{1}{\bar}$. The bottom walls of the domain are modeled as adiabatic no-slip walls. In the open-bounded simulations, the side walls are modeled as non-reflecting outflow boundaries, and in the enclosed simulations, the side walls are modeled as adiabatic no-slip walls. For the simulations with the P1 radiation model, the outlet and inlet boundaries are assumed to be perfectly non-reflecting blackbodies at the same temperature as the local gas field. The solid no-slip walls are assumed to be gray bodies with an emissivity of $\epsilon_\mrm{wall} = 0.9$ corresponding to an approximate value for white brick.

\begin{figure*}
    \centering
    \includegraphics[width=\textwidth]{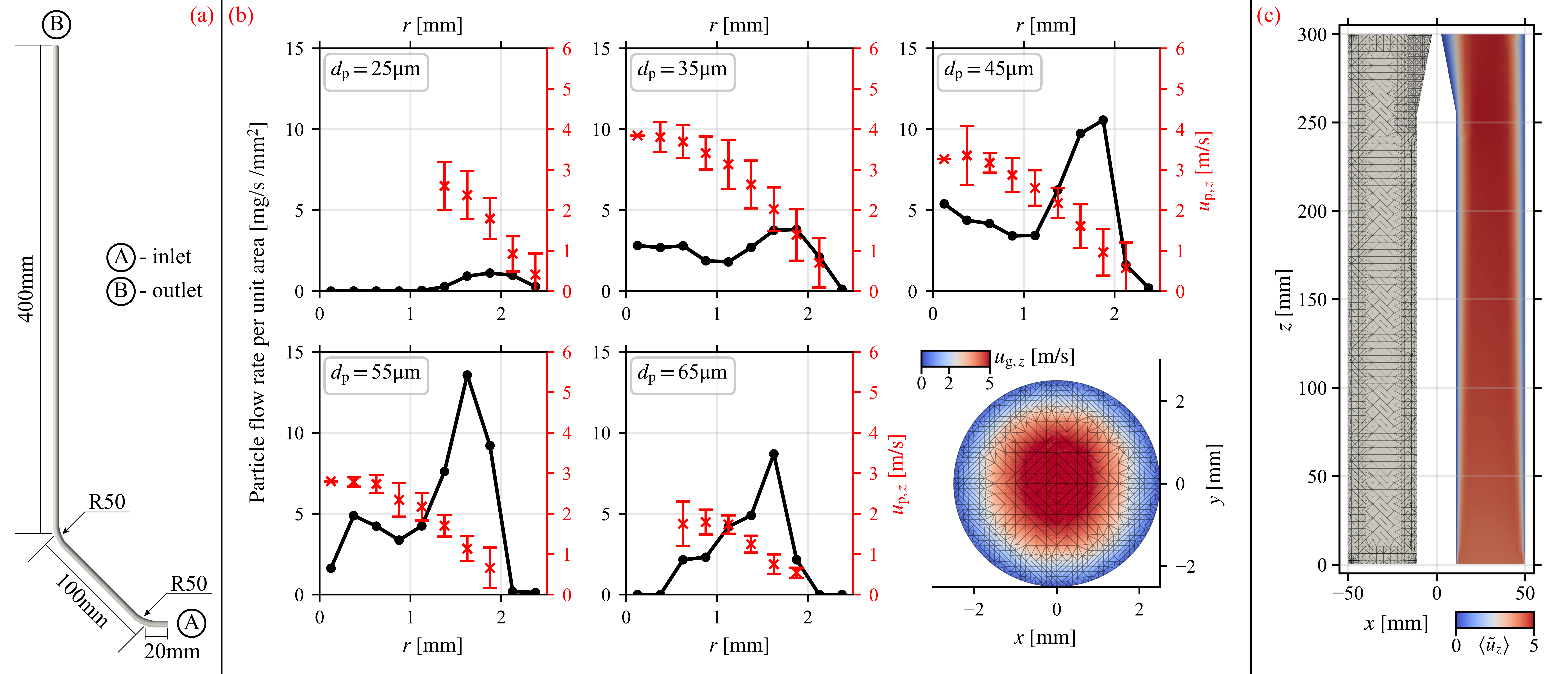}
    \caption{Schematic of the precursor simulations for the particle injection and coflow inlet. The geometric configuration of the particle injection tube is shown in subfigure (a), with per-particle flow rate per surface area and exit velocities $u_{\mrm{p},z}$, and gas velocities $u_{\mrm{g},z}$ shown in subfigure (b). The upstream region of the coflow is modeled with a turbulent HIT inlet with a uniform velocity and temperature profile and $5\%$ turbulence intensity, and the mean axial velocity $\tilde{u}_{\mrm{g},z}$ shown in subfigure (c).}
    \label{fig:inletbc}
\end{figure*}

As Hameete \cite{hameetethesis} does not provide extensive data regarding the flow and particle geometric and thermophysical properties at the injection, high-fidelity simulations are executed to provide mapping data to be used as inlet conditions. Particle entrainment in channel flows has been studied extensively in the field of fluid mechanics, and it is well-known that the radial particle concentration distribution is non-uniform and varies with particle sizes \cite{lau2016effect}. The particle injection tube, with dimensions matching the experimental reference as shown in Figure \ref{fig:inletbc}(a), is simulated as a two-way coupled Eulerian-Lagrangian DNS with the same particle distribution as in Table \ref{tab:polydispersity} to determine the particle distribution and velocities of both the carrier and the particles at the exit of the injection tube. Stochastic inter-particle collisions and elastic particle-wall collisions are considered in the simulation. Heat transfer is not considered in this precursor simulation. The resulting particle distribution and equivalent exit velocity $u_{\mrm{p},z}$ of the particles, and $u_{\mrm{g},z}$ of the gas, at the exit of the injection tube are shown in Figure \ref{fig:inletbc}(b), and are mapped as the inlet boundary condition for the particle injection in the main simulations. Due to the absence of phase distribution data, a stochastic radial velocity component following a phenomenological approach is prescribed at injection to match experimental jet spread angles. This velocity was scaled by $r/R$  and $d_\mrm{p}^{-3}$ to account for the higher inertial resistance of larger Fe particles \cite{njue2021numerical}. Here $r$ is the radial position of the particles at the particle injection inlet and $R$ is the radius of the particle injection inlet. This approach ensures that the phase distribution at the inlet is consistent with experimental snapshots. Figure \ref{fig:flamecomparison} shows the comparison of the jet spread close to the nozzle with a snapshot from the experiments of Hameete \emph{et al.} \cite{hameete2024aerosol}. The default particle injection temperature is set at $\SI{300}{\kelvin}$. The experimental setup of Hameete states that air-cooling is used to isolate the particle injection from the hot coflow, but it does not specify the exact values of particle and gas temperatures at injection. Hence, a sensitivity study on the injection temperature is performed with injection temperatures of $\SI{600}{\kelvin}$, and the coflow temperature.

The upstream region of the coflow is modeled using LES, assuming a uniform HIT upstream inlet with a uniform temperature profile. The geometry of the upstream region is modeled in reference to the experimental setup, where a honeycomb mesh is used $\SI{300}{\milli\meter}$ upstream of the injection. The turbulence intensity at the inlet of this precursor simulation is set to $5\%$ with a turbulent length scale of $L = \SI{5}{\milli\meter}$. The walls are modeled as adiabatic no-slip walls, and the outlet is modeled as a non-reflecting outflow with an ambient pressure of $p_\infty = \SI{1}{\bar}$. The filtered velocity, temperature, and turbulent kinetic energy fields from the precursor simulation of the coflow are mapped as the inlet boundary conditions for the coflow in the main simulations, with a temporal resolution of $\Delta t = \SI{1e-4}{\second}$ to capture the temporal fluctuations of the coflow. The mean velocity of the upstream region of the coflow $\langle \tilde{u}_z \rangle$ is shown in Figure \ref{fig:inletbc}(c) at the midplane of the upstream region. Air with $\tilde{Y}_\mrm{O_2}=0.233$ and $\tilde{Y}_\mrm{N_2}=0.767$ is used as the carrier phase at both inlets since the coflow is heated electrically in the experiments.

The particle distribution is polydisperse, as shown in Table \ref{tab:polydispersity}, and the Sauter diameter is maintained at $D_{3,2} = \SI{46.93}{\micro\meter}$, equivalent to the work of Hameete \emph{et al.} \cite{hameete2024aerosol}. In all the simulations, a mass injection rate of $\dot{m}_\mrm{p} = \SI{300}{\milli\gram/\second}$ is simulated. The initial Fe mass fraction of the particles is set to $Y_\mrm{Fe,0} = 0.999$ with a small initial oxide layer \cite{Paidassi1958}.

\begin{table}[h!] \footnotesize
\caption{Statistics of the particle sizes $d_\mrm{p}$ and the corresponding injection rate in terms of particles numbers $\dot{N}_\mrm{p}$, and injected mass $\dot{m}_\mrm{p}$ used in all of the simulations.}

\renewcommand{\arraystretch}{1}
\vspace{6pt}
    \centering
    \begin{tabular}{c>{\hspace{-3pt}}c>{\hspace{-3pt}}c>{\hspace{-3pt}}c}
    \hline
\rule{0pt}{1em}Bin & $d_\mrm{p}\ \mrm{[\SI{}{\micro\meter}]}$ & $\dot{N}_\mrm{p}\ \mrm{[/\SI{}{\second}]}$ & $\dot{m}_\mrm{p}\ \mrm{[\SI{}{\milli\gram/\second}]}$  \\
\hline
         1 & 25 & 165k & 10.64 \\
         2 & 35 & 253k & 44.76 \\
         3 & 45 & 250k & 94.09 \\
         4 & 55 & 144k & 98.75 \\
         5 & 65 & 46k & 51.76 \\
\hline
         $D_{3,2}$ & 46.93 & $\dot{m}_\mrm{total}$ & 300 \\
\hline
    \end{tabular}
    \label{tab:polydispersity}
\end{table}

\subsection{Simulation parameters\label{eq:setup}} \addvspace{10pt}

\begin{figure}[h]
    \centering
    \includegraphics[width=0.9\linewidth]{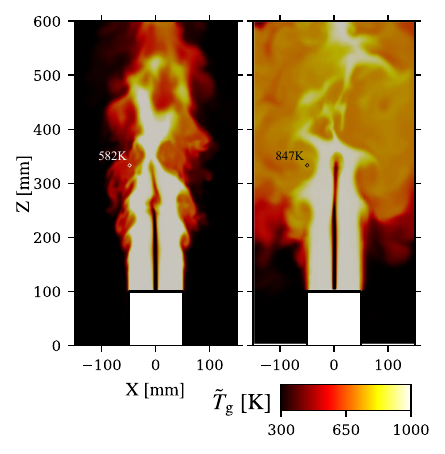}
    \caption{Mid-plane gas temperature $T_\mrm{g}$ from the precursor simulations of open flame (left) and enclosed flame (right), for a coflow temperature of $T_\mrm{coflow} = \SI{1000}{\kelvin}$.}
    \label{fig:openvsclosed}
\end{figure}

\begin{figure}[h]
    \centering
    \includegraphics[width=0.8\linewidth]{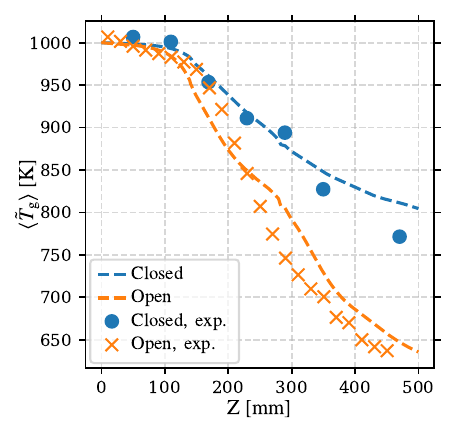}
    \caption{Comparison of time-averaged mean temperature $T_\mrm{mean}$ for open flame and enclosed flame with the experimental results from Hameete \cite{hameetethesis} at different streamline heights $z$ above the edge of the coflow.}
    \label{fig:tempvsh}
\end{figure}

The \textit{dynamicKEqn} subgrid-scale model available in \textit{OpenFOAM} is used in all the simulations with box filtering. For all the simulations, the flow field is first initialized and stabilized up to $24\tau_\mrm{f}$, where $\tau_\mrm{f}$ is the flow-through time of the coflow. Figure \ref{fig:openvsclosed} shows a comparison of the mid-plane temperature profile for the stabilized flow field of open and wall-bounded cases, indicating a difference of approximately $\SI{260}{\kelvin}$ in the gas temperature in the vicinity of the coflow. A comparison of the streamwise temperature profile with the experimental data is presented in Figure \ref{fig:tempvsh}, and good agreement with the experimental reference \cite{hameetethesis} is observed.

Particles are then injected and the simulation is executed for 14 flow-through times of the coflow to ensure convergence. The results of the oxidation degree $Y_\mrm{FeO}$ are computed at the outlet, and at various planes normal to the flow direction. 

\section{Results \& discussion \label{sec:results}} \addvspace{10pt}

In all the results discussed below, the simulations with the simplified radiation model and the wall-bounded domain are presented as the reference case unless otherwise specified. 

\subsection{Flame structure\label{sec:flamestructure}} \addvspace{10pt}

\begin{figure}[h]
    \centering
    \includegraphics[width=0.65\linewidth]{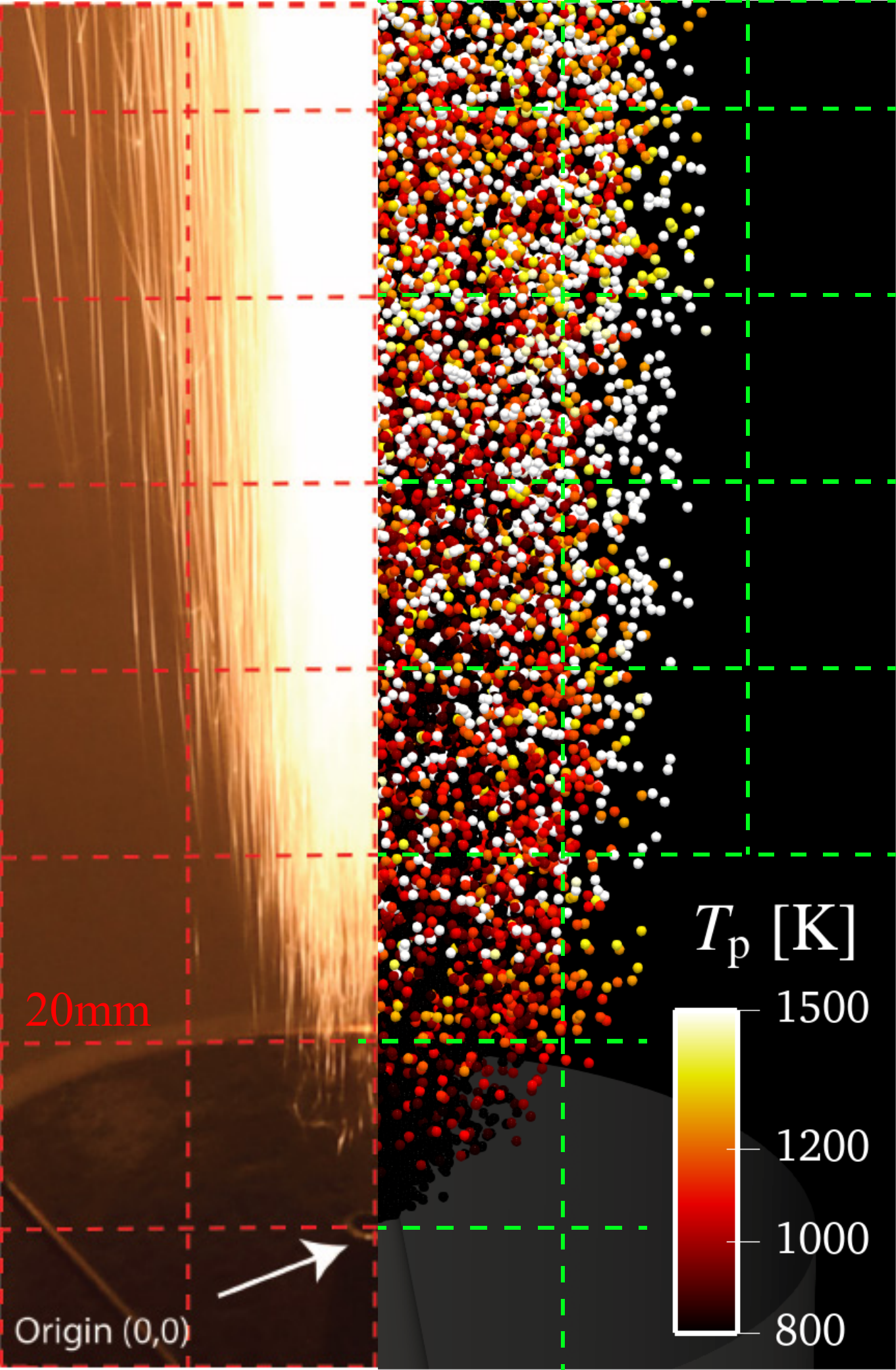}
    \vspace{10pt}
    \caption{Visualization of particles colored by temperature $T_\mrm{p}$ and comparison with the experimental snapshot adapted from Hameete \cite{hameetethesis}.}
    \label{fig:flamecomparison}
\end{figure}

Figure \ref{fig:flamecomparison} shows a visual comparison of the jet dispersion through the particle positions and temperature $T_\mrm{p}$, along with a snapshot from experiments \cite{hameetethesis}.

\begin{figure}[h]
    \centering
    \includegraphics[width=0.95\linewidth]{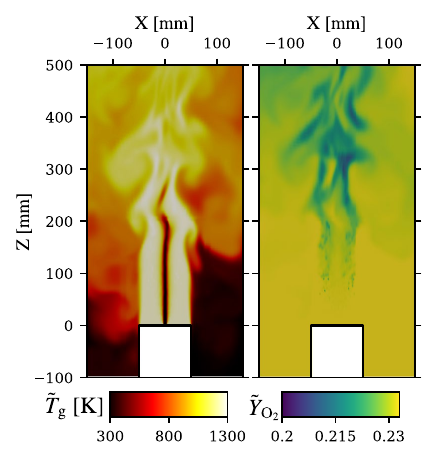}
    \caption{Visualization of the midplane gas-phase temperature $\tilde{T}_\mrm{g}$ and oxygen mass fraction $\tilde{Y}_\mrm{O_2}$ of an enclosed iron powder flame in hot coflow at $T_\mrm{coflow} = \SI{1200}{\kelvin}$.}
    \label{fig:midplanegasflame}
\end{figure}

Figure \ref{fig:midplanegasflame} shows the mid-plane carrier phase temperature $T_\mrm{g}$ and oxygen mass fraction $Y_\mrm{O_2}$ at $T_\mrm{coflow}=\SI{1200}{\kelvin}$. Since the mass-loading with respect to the coflow results in a lean mixture, only a minor depletion in oxygen concentration is observed. Ignition occurs at the circumference of the jet, where the particles are in contact with the hot coflow, and the flame propagates inwards. However, at a height of approximately $z=\SI{200}{\milli\meter}$ above the injection nozzle, the coflow, and subsequently the particle jet, breaks up. While the onset of ignition is indiscernible from the temperature field $\tilde{T}_\mrm{g}$, the oxygen mass fraction field $Y_\mrm{O_2}$ reveals the onset of ignition at a height of approximately $z=\SI{100}{\milli\meter}$, where the oxygen mass fraction starts to deplete in the vicinity of the jet circumference. Supplementary material \texttt{gastemp.mp4} and \texttt{gasyo2.mp4} show the temporal evolution of the mid-plane gas temperature and oxygen mass fraction, respectively, for the reference case at $T_\mrm{coflow}=\SI{1150}{\kelvin}$.

\begin{figure}[h]
    \centering
    \includegraphics[width=0.9\linewidth]{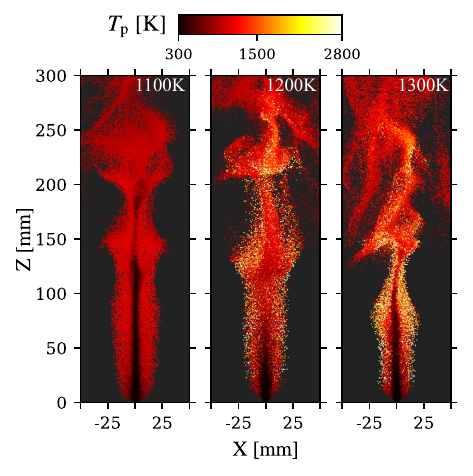}
    \caption{Visualization of particle temperature $T_\mrm{p}$ clipped mid-plane at $T_\mrm{coflow}=\SI{1100}{\kelvin}$, $\SI{1200}{\kelvin}$ and $\SI{1300}{\kelvin}$ for enclosed iron powder flame. No ignition is observed at $T_\mrm{coflow}=\SI{1100}{\kelvin}$, ignition is observed at $T_\mrm{coflow}=\SI{1200}{\kelvin}$ and at $T_\mrm{coflow}=\SI{1300}{\kelvin}$, although the flame structure is different from that at $T_\mrm{coflow}=\SI{1200}{\kelvin}$, with a more uniform ignition across the jet and earlier completion of oxidation.}
    \label{fig:midplaneptclflame}
\end{figure}

Visualization of the particle temperature $T_\mrm{p}$ at various coflow temperatures $T_\mrm{coflow}$ is shown in Figure \ref{fig:midplaneptclflame}. The temporal evolution of the particle jet at $T_\mrm{coflow} = \SI{1150}{\kelvin}$ is also visualized in the supplementary material \texttt{ptcltemp.mp4}. At $T_\mrm{coflow} = \SI{1100}{\kelvin}$, no ignition is observed. The particles heat up and homogenize with the flow. At $T_\mrm{coflow} = \SI{1200}{\kelvin}$, onset of ignition is observed at the circumference of the jet. However, ignition of the core of the particle jet does not occur until the break-up of the particle jet. Subsequently, the particle jet evolves into multiple clouds of particles, and secondary ignition progresses in these clouds, as evident also from the depletion of $\mrm{O_2}$ in these clouds as in the supplementary material \texttt{gasyo2.mp4}. The reader is referred to the mentioned supplementary material for a detailed visualization. At $T_\mrm{coflow} = \SI{1300}{\kelvin}$, the flame structure shows earlier ignition of the particles across the jet, and oxidation is completed at a height of approximately $z=\SI{150}{\milli\meter}$ above the injection nozzle, with no secondary ignition in broken-up particle clouds. This indicates that the ignition progress can be different at different coflow temperatures, and the transition between these mechanisms can be quite sensitive to the coflow temperature and the flow physics.

\subsection{Ignition characteristics} \addvspace{10pt}

\begin{figure}[h]
    \centering
    \includegraphics[width=0.9\linewidth]{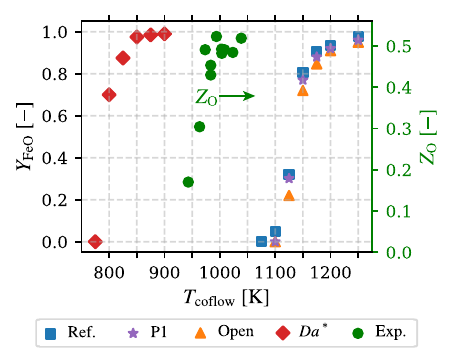}
    \caption{Oxidation degree $Y_\mrm{FeO}$ of the particles crossing the exit plane ($z=\SI{0.5}{\meter}$ above injection) at various coflow temperatures $T_\mrm{coflow}$ of the reference case (Ref.) of an enclosed flame with the simplified radiation model and the oxide layer model, compared with the results from using the P1 model for radiation (P1), open-walled case (Open), the first order model for reaction kinetics ($Da^*$) and the experimental results of oxidation degree $Z_O$ by Hameete \cite{hameetethesis}.}
    \label{fig:oxidation}
\end{figure}

Particles crossing an $xy$-plane at various streamwise positions $z$ are tracked during the simulations. The mean oxidation degree $Y_\mrm{FeO}$ is then computed from the mass of the tracked particles compared to their initial mass. The comparison of results, as shown in Figure \ref{fig:oxidation}, is conducted at a plane height of $z=\SI{0.5}{\m}$ above the injection nozzle.

The first-order model completely overpredicts ignition owing to the low $T_\mrm{ign}$ and exhibits the earliest ignition at $T_\mrm{coflow}=\SI{800}{\kelvin}$ with complete oxidation at $T_\mrm{coflow} = \SI{900}{\kelvin}$. On the other hand, the oxide-layer model, in general, underpredicts ignition and shows complete oxidation only at $T_\mrm{coflow} =\SI{1250}{\kelvin}$. Compared to the results of Hameete \emph{et al.} \cite{hameete2024aerosol}, a deviation of approximately $\SI{150}{\kelvin}$ is noted. 

Comparing the two modes, open and enclosed flame, higher oxidation degrees are seen in wall-bounded flames. This result is as expected since the gas temperature surrounding the coflow is higher, as shown in Figure \ref{fig:openvsclosed}, contributing to less net heat transfer away from the particle. Comparing the results of the simulations with the simplified radiation model to those with the P1 radiation model, it is observed that the P1 model results in a slightly lower oxidation degree, although the difference is insignificant, inferring that radiation is not an important phenomenon that influences ignition in such flames.

\subsubsection{Discrepancy in oxidation degree}\addvspace{10pt}

\begin{figure}[h]
    \centering
    \includegraphics[width=0.9\linewidth]{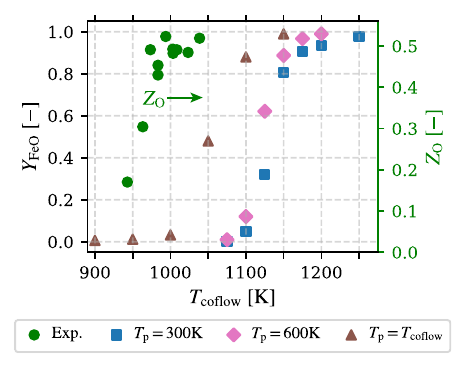}
    \caption{Oxidation degree $Y_\mrm{FeO}$ of the particles crossing the exit plane ($z=\SI{0.5}{\meter}$ above injection) at various particle injection temperatures $T_\mrm{p}$. }
    \label{fig:oxidationTp}
\end{figure}

The discrepancy in the oxidation degree between the simulations and the experiments can be attributed to several factors. The model parameters used in this work are based on prior studies and may not be fully representative of the particle properties used in the experiments. The following are unknowns from the experimental setup that could contribute to the discrepancy in the oxidation degree:

\begin{enumerate}[leftmargin=*]
    \item Efficiency of air-cooling of particle injection tube
\end{enumerate} \vspace{-3pt}
The efficiency of the air-cooling in isolating the particle injection tube from the coflow is unknown, and the particle injection temperature could be higher than $\SI{300}{\kelvin}$ at injection. An injection temperature of $\SI{300}{\kelvin}$ assumes 100\% efficiency of air-cooling, and in reality, the particles may be injected at a higher temperature which may further improve the oxidation degree. Figure \ref{fig:oxidationTp} shows the oxidation degree at various particle injection temperatures $T_\mrm{p}$. However, even when the particles are injected at the same temperature as the coflow, the oxidation degree exhibits a discrepancy with the experimental results. Hence, the particle injection temperature alone cannot explain the discrepancy in the oxidation degree.

\begin{enumerate}[leftmargin=*]
    \setcounter{enumi}{1}
    \item Surface morphology of the particles
\end{enumerate}\vspace{-3pt}
The particles are assumed to be spherical and non-porous in the simulations, while the actual particles used in the experiments might have a more complex morphology, which can affect the ignition characteristics. The particle morphology can affect the surface area available for reaction and heat transfer, which can influence the ignition behavior. Cen \emph{et al.} \cite{cen2025detailed} show that the ignition of porous or irregular particles is substantially different from that of solid particles. Hameete \cite{hameetethesis} reports that the particles used in the experiments are irregularly shaped, which could contribute to the discrepancy in the oxidation degree. A recent study by St. Germain \emph{et al.} \cite{st2026iron} has shown that the irregular morphology of iron particles can lower the critical temperature compared to spherical particles by up to $\SI{150}{\kelvin}$. This can be a contributing factor to the discrepancy in the oxidation degree observed in the present work.

\subsubsection{What causes partial oxidation?} \addvspace{10pt}
The case of a wall-bounded flame at $T_\mrm{coflow}=\SI{1150}{\kelvin}$ with the simplified radiation model shows an overall oxidation degree of $Y_\mrm{FeO} = 0.805$, indicating partial oxidation, and is further analyzed. 

\begin{figure}[h]
    \centering
    \includegraphics[width=0.95\linewidth]{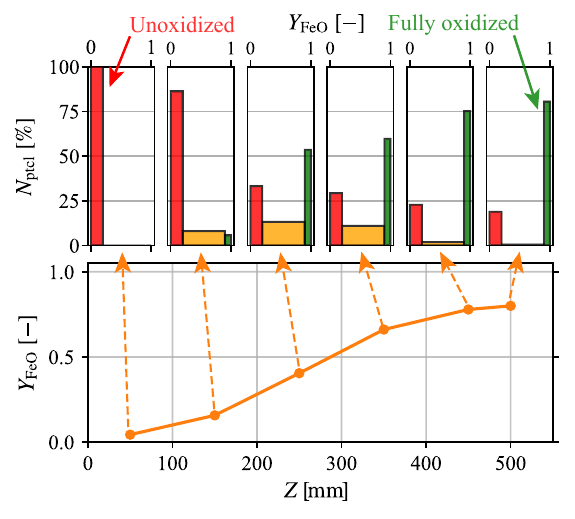}
    \caption{Histogram of $Y_\mrm{FeO}$ of particles crossing an $xy$-plane at various streamline heights $z$ showing unignited (red), actively reacting (orange) and completely oxidized (green). The bottom plot shows the overall oxidation degree $Y_\mrm{FeO}$ at various streamline heights $z$. These results are at a coflow temperature of $T_\mrm{coflow}=\SI{1150}{\kelvin}$ for an enclosed flame with the simplified radiation model.}
    \label{fig:oxidationheight}
\end{figure}

Figure \ref{fig:oxidationheight} shows the histogram of the mass fraction of $\mrm{FeO}$ of particles tracked at different streamwise heights for the discussed case. Notably, a binary distribution of $Y_\mrm{FeO}$ is observed at larger $z$, where there are no partially oxidized particles--either unignited particles with $Y_\mrm{FeO} < 0.2$ or completely oxidized particles with $Y_\mrm{FeO} \approx 1$, indicating that all ignited particles oxidize completely. Hence, at a certain height above the burner, the overall oxidation degree reflects the probability of ignition. This further validates the observations by van Rooij \emph{et al.} \cite{niekvanrooijthesis,van20260}, and other \emph{a posteriori} analyses from large-scale burners, that the particle distribution at the exit of the burner is binary, with a fraction of unignited irregular $\mrm{Fe}$ particles with an oxide layer and the remainder being completely oxidized spherical $\mrm{Fe_3O_4}$ particles. The evolution of $\mrm{Fe}$ mass fraction $Y_\mrm{Fe}$ of the particles is shown in the supplementary material \texttt{ptclyfe.mp4} , which also indicates the progression of ignition and the binary distribution.
\begin{figure}[h]
    \centering
    \includegraphics[width=\linewidth]{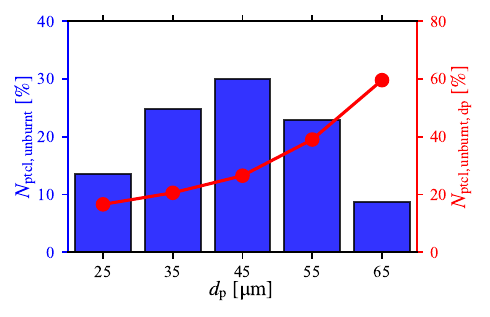}
    \caption{Distribution of unburnt particles among all exiting particles $N_\mathrm{ptcl,unburnt}$ (blue histogram) and the ratio of unburnt particles normalized per particle size $N_\mrm{ptcl,unburnt,d_\mrm{p}}$  (red line). These results are at a coflow temperature of $T_\mrm{coflow}=\SI{1150}{\kelvin}$ for an open flame with the simplified radiation model.}
    \label{fig:sizeanalysis}
\end{figure}

Analyzing the oxidation statistics based on the particle diameters, as shown in Figure \ref{fig:sizeanalysis}, indicates that the majority of unburnt particles (blue histograms in Figure \ref{fig:sizeanalysis}) are of size $d_\mrm{p} = \SI{45}{\micro\meter}$. However, this is merely because $\SI{45}{\micro\meter}$-sized particles are more numerous (see Table \ref{tab:polydispersity}), and normalizing with the total number of particles of the same size (red line in Figure \ref{fig:sizeanalysis}) shows that the failure of ignition is slightly more prevalent for larger particles. This is hypothesized to be an effect of slower heating-up of larger particles, which have higher thermal capacity and hence require a longer time to ignite relative to the residence time in the hot coflow. Mi \emph{et al.} \cite{mi2025theoretical} showed that slower heating of the particle increases the critical temperature required for thermal runaway due to the buildup of the oxide layer. Further analysis into the heating history of the particles to ignition is required to confirm this hypothesis.

\section{Conclusion \label{sec:conclusion}} \addvspace{10pt} 

The jet-in-hot-coflow iron powder burner designed by Hameete \emph{et al.}\cite{hameete2024aerosol} is computationally modeled using LES with an Eulerian-Lagrangian approach. The flow and particle properties are chosen to resemble the experiments conducted by Hameete \emph{et al.}. 

At lower temperatures $\SI{1125}{\kelvin} < T_\mrm{coflow} < \SI{1200}{\kelvin}$, onset of ignition is observed at the circumference of the jet and is further aided by the break-up of the particle jet, which results in secondary ignition post break-up of the particle jet.  At higher temperatures $T_\mrm{coflow} > \SI{1200}{\kelvin}$, oxidation is mostly complete before the jet break-up.

Ignition is observed at $T_\mrm{coflow} \approx \SI{1125}{\kelvin}$ for the simulated cases with the oxide layer model by Mi \emph{et al.} \cite{Mi2022}, with complete oxidation at $T_\mrm{coflow}\approx\SI{1250}{\kelvin}$, which is higher than the earliest ignition temperature of $\SI{950}{\kelvin}$ and the complete oxidation temperature of $\SI{1050}{\kelvin}$ observed in the experiments. The oxidation degree was slightly higher in the enclosed flame compared to the open flame, and the P1 model showed a slightly lower oxidation degree compared to the simplified radiation model. The first-order model formulated by Hazenberg and van Oijen \cite{Hazenberg} overpredicts the oxidation degree, while the oxide-layer model underpredicts the oxidation degree compared to the experimental results.

Analysis of the particles crossing planes at various streamline heights $z$ shows the progression of global oxidation. At higher $z$, the particle distribution is binary, consisting of either fully oxidized particles or totally unignited particles, thereby explaining partial oxidation. The analysis also reveals that the failure of ignition is prevalent in larger particles, possibly due to their inadequate residence time in the coflow and slower heating.

The discrepancy in the oxidation degree between the simulations and the experiments can be attributed to several factors, including the assumptions made in this work and the unknowns in the experimental setup. The assumptions regarding the particle injection temperature and the particle morphology are expected to have a significant effect on the ignition characteristics and, hence, the oxidation degree. The lack of detailed injection boundary conditions of the JHC burner and the oxidation model of irregular, rough-surface iron particles requires future efforts to resolve the observed discrepancies in the results.

\acknowledgement{CRediT authorship contribution statement} \addvspace{10pt}

{\bf S.H.}: Conceptualization, Methodology, Software, Validation, Formal analysis, Investigation, Resources, Data curation, Writing - Original Draft, Visualization {\bf X.C.M.}: Conceptualization, Methodology, Investigation, Resources, Writing - Review \& Editing, Supervision, Project administration, Funding acquisition

\acknowledgement{Declaration of competing interest} \addvspace{10pt}

The authors declare that they have no known competing financial interests or personal relationships that could have appeared to influence the work reported in this paper.

\acknowledgement{Acknowledgments} \addvspace{10pt}

This project has received funding from the EIRES Start-Up Package (CRT STA MS-FeComb). This work used the Dutch national e-infrastructure with the support of the SURF Cooperative, under grant number 2024.003 and grant number EINF-16502. The authors would like to thank Drs. Giel Ramaekers, Thijs Hazenberg, and Bharat Bhatia for their advice on the computational setup.

\footnotesize
\baselineskip 9pt

\clearpage
\thispagestyle{empty}
\bibliographystyle{proci}
\bibliography{references}


\newpage

\small
\baselineskip 10pt


\end{document}